\documentclass[twocolumn,pra,aps,showpacs]{revtex4-1}
\usepackage{xspace,amsmath,amsfonts,amsthm,amssymb,amsbsy,graphicx,color}

\begin{document}

\title{Frequency Conversion: Side-band cooling, state-swapping, and coherent control of mechanical resonators}

\author{Kurt Jacobs$^1$, Hendra I. Nurdin$^2$, Frederick W. Strauch$^3$, and Matthew James$^2$} 

\affiliation{$^1$Department of Physics, University of Massachusetts at Boston,
Boston, MA 02125, USA \\
$^2$School of Engineering, The Australian National University, 
Canberra, ACT 0200, Australia \\
$^3$Department of Physics, Williams College, Williamstown, MA 01267
}

\begin{abstract} 
Sideband cooling is a technique that potentially allows mechanical resonators to be prepared in their ground states, important for future applications in quantum technologies. Tian has recently shown that side-band cooling can be implemented by modulating the coupling between a nano-resonator and a superconducting oscillator, a process of frequency conversion [L. Tian, PRB \textbf{79}, 193407 (2009)]. While side-band cooling is usually treated in the steady-state regime, the effective resonant coupling will also generate near perfect state-swapping from the superconductor to the mechanical resonator. We perform numerical simulations of this system, examining the ground-state cooling achieved by the state-swapping. Further, we show that the superconducting oscillator can be used to control the amplitude and phase of the resonator, while simultaneously cooling it, and thus act as a coherent ``quantum feedback controller''. 
\end{abstract}


\maketitle 

Nano-mechanical resonators can now be built with high frequencies and high quality factors~\cite{Regal08, LaHaye09, Rocheleau10} and interfaced with mesoscopic superconducting circuits~\cite{Majer07, Houck07}. In order to observe and exploit the coherent quantum properties of these devices, one must first be able to cool them to their ground state. Ground-state cooling is achieved when the average number of phonons in the resonator, $\langle n \rangle$, (also called the ``average occupation number'') is much less than unity. Feedback control via measurements could in theory be used to perform such cooling~\cite{Hopkins03}, but this approach is not yet feasible with current technology. Two classes of methods have been devised to date to achieve ground-state cooling without using measurements. The first involves periodic coupling to a superconducting qubit or qutrit~\cite{Zhang05, You08}. The second consists of various ways to realize \textit{side-band} cooling~\cite{Wineland79,  Wilson-Rae04, Tian09}, also known as ``radiation pressure'' cooling or ``dynamical back-action'' cooling. Recently Tian~\cite{Tian09} proposed performing sideband cooling by using a linear coupling between a nano-resonator and a ``auxiliary'' superconducting L-C oscillator, and modulating this coupling at the difference frequency between the two resonators. This formulation is especially nice, because it shows  that the cooling comes from the fact that the frequency conversion makes the auxiliary oscillator look to the mechanical resonator as if it is much colder, even though it is really at the same ambient (background) temperature. 

In a nutshell, sideband cooling viewed as frequency conversion can be described as follows. If one couples a low-frequency resonator (which we will call the \textit{target}) to a high-frequency resonator (the \textit{auxiliary}) via a linear coupling modulated at the resonators' difference frequency, then the two resonators appear to each other as if they are on-resonance. Since both resonators are at the ambient temperature, the auxiliary has a lower occupation number (is closer to its ground state), and therefore looks to the target as if it is at a lower temperature. If we now ensure that the auxiliary has a \textit{much faster damping rate} than the target, energy that flows from the target to the auxiliary is quickly lost to the bath. The result is that the target is cooled close to the effective temperature of the auxiliary. This is the cooling method in a nutshell, minus a couple of important additional details that we will discuss below. The above scenario is perfectly suited for cooling a nano-resonator by coupling it to a superconducting L-C oscillator~\cite{Tian09} or ``stripline" resonator~\cite{Regal08, Majer07, Houck07}. This is because these superconducting resonators have both much higher frequencies and much higher damping rates than their mechanical cousins. 

The caveat to the above description is that the modulated coupling only generates perfect transfer of energy quanta between the two resonators to the extent that the rotating wave-approximation (RWA) is valid. This is the case so long as the coupling rate between the oscillators, as well as their damping rates, are small compared to the frequencies of both oscillators. If the rotating wave approximation is not valid, then the coupling also generates quanta in both oscillators, leading to heating. This is why side-band cooling only provides good ground-state preparation in the resolved-sideband limit, in which the damping rate of the auxiliary is much smaller than the frequency of the mechanical resonator.  

In addition to providing sideband cooling, a resonant linear coupling between two oscillators has the remarkable property that it generates a perfect state-swap between the oscillators for an evolution time of $\tau = \pi/(2g)$, where $g$ is the interaction rate (defined precisely below)~\cite{Tian08}. Since the auxiliary oscillator is initially in its ground state (at the ambient temperature), a single state-swap will prepare the target in its ground state, thus providing a second way to achieve ground-state cooling. We examine both methods of cooling here, via numerical simulations. Note that this technique of cooling by swapping the oscillator states is closely related to the cooling method using an auxiliary qubit suggested by Zhang \textit{et al.}~\cite{Zhang05}. The present method is much more efficient, however, as the auxiliary oscillator can absorb all the energy in a single interaction time $\tau$, while on this time-scale an auxiliary qubit only absorbs a single quanta.

In the second part of our analysis, we show that the frequency conversion coupling can be used to control the coherent state of the resonator, while simultaneously providing sideband cooling. In this way, the L-C oscillator reduces the noise in the target system as well as controlling it in real time. This is precisely the action of a feedback control loop --- note that the noise that is eliminated by the cooling process does not have to be thermal in origin, it is equally effective at extracting noise from any source. In the usual feedback loop, measurements are made on the system, and actions taken based on these measurements to reduce the noise and effect control. The frequency-conversion coupling with the L-C oscillator is therefore an example of a \textit{coherent} feedback control loop~\cite{SL00, YK03a, Gough10, James08b, Nurdin09}, in which the information flows directly from the target system, to another quantum system that acts as a controller, without the use of measurements. There is therefore a fundamental connection between sideband cooling, state-swapping, and coherent quantum feedback control. 

Expressions for the average occupation number achievable by sideband cooling in the steady-state have been derived in~\cite{Marquardt07, Tian09}. We note that one can also derive a very simple approximate expression for the steady-state cooling, and we present this now for comparison with the numerical results presented below. In what follows we will denote the frequencies of the target and auxiliary systems as $\omega$ and $\Omega$, respectively, and their damping rates as $\gamma$ and $\kappa$. We will also denote the coupling rate between the two systems as $g$, and assume that the average occupation number of the auxiliary is negligible at the ambient temperature (this is true for superconducting resonators with $\Omega/(2\pi) \geq 5~\mbox{GHz}$ in dilution refrigerators). If we take the coupling constant $g$ to be at least as large as the auxiliary damping rate $\kappa$, then it seems reasonable that the energy flow rate out of the target will be limited by, and thus approximately equal to, $\kappa$. In this case equating the heating and cooling rates, one obtains an estimate for the steady-state occupation number:   
\begin{equation}
     \langle n \rangle_{\mbox{\scriptsize c}}  \approx  \left[ \gamma/(\gamma + \kappa) \right]  n_T , 
       \label{eq::rsb} 
\end{equation} 
where $n_T$ is the occupation number at the ambient temperature $T$. An example for realistic parameters would be a $20~\mbox{MHz}$ nano-resonator with quality factor of $10^5$ ($\gamma/2\pi = 200~\mbox{Hz}$) coupled to an L-C oscillator with a frequency of $5~\mbox{GHz}$ and a damping rate of $\kappa = 10^6~\mbox{s}^{-1}$~\cite{Frunzo05}. Choosing a coupling rate of $g = 2 \times 10^{6}~\mbox{s}^{-1}$ gives $\langle n \rangle_{\mbox{\scriptsize c}} \approx 1\times 10^{-3} n_T$. 

\textit{Numerical results for cooling:} We now treat the frequency conversion, steady-state cooling, and state-swapping in more detail. We perform numerical simulations to check Eq.(\ref{eq::rsb}), and the limiting effect of the coupling. We consider the following linear coupling between a nano-resonator and a superconducting resonator:  
\begin{equation}
    H = \hbar \omega a^\dagger a +  \hbar \lambda (a + a^\dagger)(b + b^\dagger ) + \hbar \Omega b^\dagger b , 
\end{equation} 
in which $a$ and $b$ are the annihilation operators for the nano-resonator and stripline, respectively. Here $\lambda$ is the strength of the coupling between the two oscillators. Tian has shown explicitly how to obtain a linear coupling between a nano-resonator and an L-C resonator~\cite{Tian09}. Alternatively a mechanical resonator can be coupled to a superconducting ``stripline'' resonator via the nonlinear interaction $H_{\mbox{\scriptsize int}} \propto b^\dagger b (a + a^\dagger)$, or to an optical cavity via the same interaction (in this case it is referred to as ``radiation-pressure''). By driving the stripline one can obtain from this an effective linear interaction, and this is the usual context in which radiation-pressure cooling has been analyzed in the past. 

To achieve the frequency conversion we now modulate the coupling strength at the difference frequency $\Delta = \Omega - \omega$, so that $\lambda = g \cos(\Delta t)$. Writing the Hamiltonian in the interaction picture, the result is 
\begin{equation}
    H_{\mbox{\scriptsize int}} = \hbar g (a b^\dagger + a^\dagger b) +  \hbar g( a b e^{-i2\omega t}  + a^\dagger b^\dagger e^{i2\omega t}  ) ,  
    \label{eq::Hint} 
\end{equation} 
where we have dropped all terms oscillating at $\Omega$, since this frequency is much higher than all the other dynamical timescales. 

The time-independent part of the interaction Hamiltonian in Eq.(\ref{eq::Hint}) describes the resonant exchange of energy quanta between the two oscillators. This is what we need for cooling, state-swapping, and coherent control. The time-dependent terms will interfere with all these processes. In the usual picture of side-band cooling, the time-dependent terms are  viewed as generating a heating rate~\cite{Tian09, Wilson-Rae07}. But the frequency conversion picture we have here does not afford us with a simple interpretation. The rotating wave approximation (RWA), which allows us to drop the time-dependent terms, requires that $g \ll \omega$. In this case the coupling between the two oscillators is perturbative, and, in particular, has a minimal effect on the ground state of the target. Since $g$ determines the rate at which energy is transferred out of the target, this ultimately places a limit on the rate of the cooling. Nevertheless, since nano-resonators have frenqecies in the 10's to 100's of MHz, cooling rates can still be on the order of $10^6~\mbox{s}^{-1}$. 

We model the thermal damping of both oscillators using the quantum-optical master equation. This is an accurate model so long as the quality factor of each oscillator is much greater than unity. For a single oscillator with annihilation operator $c$, this master equation is~\cite{WMbook}
\begin{eqnarray}
   \dot{\rho} & = &   - (\Gamma/2) (n_T + 1)( \{ c^\dagger c, \rho \}_+ - 2 c \rho c^\dagger  )  \nonumber \\
      &  & - (\Gamma/2) n_T ( \{ cc^\dagger , \rho \}_+ - 2  c^\dagger \rho c ) , 
      \label{eq::me} 
\end{eqnarray} 
where $\rho$ is the density matrix,  $\Gamma$ is the damping rate, and $c$ is the annihilation operator for the oscillator. The constant $n_T$ is the average occupation number of the oscillator at the ambient temperature, $T$, and is given by $n_T =  1/[ \exp(-\hbar\nu/kT) - 1]$, where $k$ is Boltzmann's constant and $\nu$ is the angular frequency of the oscillator. Both oscillators are subject to this master equation, with $\nu$ and $\Gamma$ replaced with the appropriate quantities. 

We now simulate the evolution of the target and auxiliary, under the interaction given by Eq.(\ref{eq::Hint}), and the thermal damping of both oscillators. To perform this simulation we use the Monte Carlo wave-function method~\cite{WM93}, combined with a Milstein stochastic integrator~\cite{JacobsSP}. We find that 4096 trajectories are sufficient to obtain accurate results. We are somewhat restricted in the size of the Fock space we can use for the simulation, but this is not especially important since it is the cooling \textit{factor} that we wish to confirm. Previous results on side-band cooling, consistent with the simple expression derived above, confirm that the cooling factor is largely independent of the ambient temperature. 

\begin{figure}
\leavevmode\includegraphics[width=1\hsize]{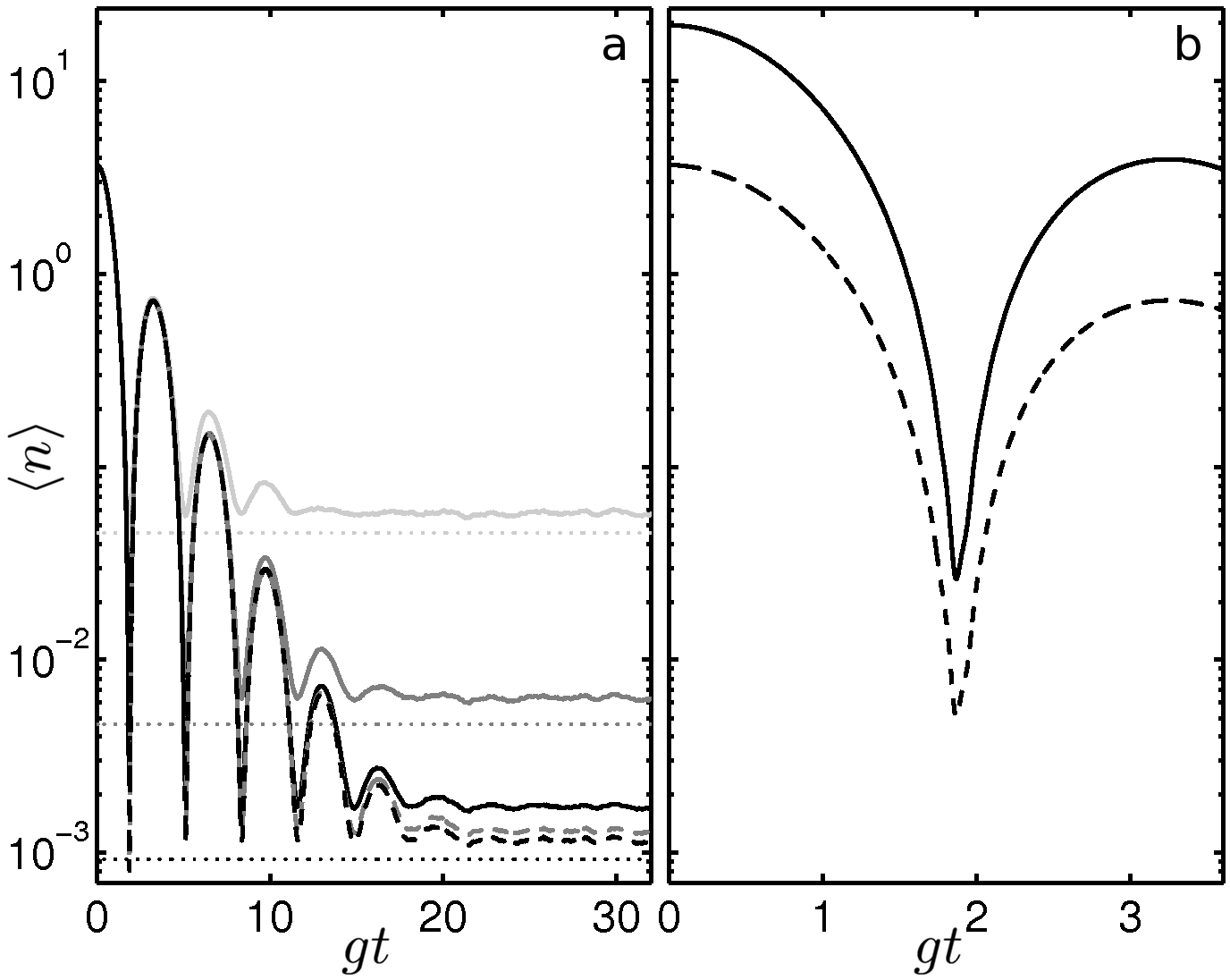}
\caption{(a) A simulation of cooling a (target) nano-resonator with frequency $\omega$, using a frequency-conversion (modulated) coupling to an (auxiliary) L-C oscillator. The initial thermal state of the target has occupation $\langle n \rangle = 3.68$. The solid lines have coupling $g = \omega/(20\pi) = 2\kappa$, where $\kappa$ is the damping rate of the auxiliary. The $Q$ values are $10^4$ (light grey), $10^5$ (medium grey), and $5\times 10^5$ (black). The dashed lines have the same damping rate for the target as the black solid line, but now with $\omega/g = 40\pi$ (grey), and $\omega/g \rightarrow\infty$ (black). The straight dashed lines are the simple steady-state estimates provided by Eq.(\ref{eq::rsb}) for the three different damping rates. \\ 
 (b) A plot showing the cooling achieved by a single state-swap. The parameters for the system are those used for the medium-grey solid line in Fig.~\ref{fig1}a. The dashed-line, for which the initial occupation number is $\langle n \rangle = 3.68$, is merely a blow-up of the medium-grey solid line in (b). The solid line has $\langle n \rangle = 20$.} 
\label{fig1}
\end{figure} 

For our first set of simulations we choose an initial thermal state for the target with occupation number $\langle n \rangle = 3.68$. This corresponds, for example, to a 100 MHz resonator at 20 mK, and can be simulated with 32 Fock-states for each resonator. In Fig.~\ref{fig1}a (solid lines) we plot the evolution of $\langle n \rangle$ for the target for a $20~\mbox{MHz}$ resonator with $g = 2 \kappa = 2~\mbox{MHz}$, and three values of the target damping rate: $\gamma/2\pi = 2~\mbox{KHz}$, $200~\mbox{Hz}$, and $40~\mbox{Hz}$ (corresponding to quality factors $Q = 10^4$, $10^5$, and $5\times 10^5$). We also plot the steady-state value for $\langle n\rangle_{\mbox{\scriptsize c}}$ as predicted by the simple formula in Eq.(\ref{eq::rsb}) (straight dashed lines). Here we see that the frequency-conversion interaction swaps the phonons between the nanomechanical resonator (the target) and the auxiliary oscillator.  The auxiliary's large damping rate cools the system until the swapping ceases. We see that for $Q = 10^4$ the steady-state value is within $30\%$ of the simple estimate. The efficacy of the estimate appears to degrade as $Q$ is reduced, but this is not the case --- as we increase $Q$, and thus the cooling factor, we are instead seeing the limiting effect of the size of $g/\omega$ (the failure of the rotating wave-approximation). To show this, we also plot $\langle n(t) \rangle$ with the time-dependent part of the interaction dropped (the black-dashed curve), and we see that the resulting steady-state is once again within $30\%$ of that predicted by Eq.(\ref{eq::rsb}). This shows us that our simple formula consistently overestimates the cooling by a factor of $\approx 1.3$, so long as $g/\omega$ is sufficiently small. From these plots we also see that the finite value of $g/\omega$ does not impose a fixed limit on the achievable temperature, but as has already been noted in previous analyses, acts like an additional heating rate~\cite{Tian09, Wilson-Rae07}. 

We now examine the state-swapping induced by frequency conversion, which is already evident in Fig.~\ref{fig1}a. Our numerical simulations show that the damping of the auxiliary, $\kappa$, has little effect on the fidelity of the swap, even when it is the same order-of-magnitude as the interaction rate. It does, however, shift the time at which the swap occurs. For example, for $\kappa = g/2$ the swap time is $1.177\pi/(2g)$. Note that while cooling in the steady state requires $\kappa \gg \gamma$, ground-state preparation using a state-swap does not. Thus if the ratio $\kappa/\gamma$ is low enough, the state-swap will achieve a lower temperature than side-band cooling. The degree to which the state-swapping can prepare the resonator in the ground state (that is, how cold it can cool) in this case depends upon the efficacy of the rotating wave approximation ($g \ll \omega$) and the heating of the target that takes place during the swap. The latter is set by the ratio of $g$ to the heating rate $\gamma n_T$. 

In Fig.~\ref{fig1}b we show the cooling achieved by a state-swap for $g = \omega/(20\pi) = 2\kappa$, $Q = 10^5$, and two values for the temperature of the target resonator, $ n_T = 3.68$ and $20$. For these values of the temperature we obtain cooling factors of $725$ and $758$, respectively, showing that the cooling factor varies only slowly with the ambient temperature. By increasing the ratios $\omega/g$ and $g/\gamma$ we can obtain higher cooling factors. An example of this is shown by the dashed-lines in Fig.~\ref{fig1}a, in which the first minimum of $\langle n \rangle$ is less than $10^{-3}$.  

\textit{Coherent feedback control:} To control the coherent state of the target --- that is, control its phase and amplitude --- first consider when the target is at zero temperature. In this case, driving the auxiliary oscillator with a classical input signal will in turn provide a coherent drive for the target via the coupling with the auxiliary. The resulting steady-state phase and amplitude of the target is thus determined by the drive on the auxiliary. Since the thermal noise affecting the target is uncorrelated with the drive, the resulting steady-state of the target is simply the cooled thermal state, displaced in phase space to the amplitude and phase dictated by the drive. The auxiliary thus simultaneously cools and controls the target. 

The coherent control is easily analyzed by using the Heisenberg equations of motion for the coupled oscillators. To do this we make the RWA, because while the motion \textit{can} be exactly solved if we keep the time-dependent terms in the Hamiltonian (Eq.(\ref{eq::Hint})), the resulting expressions are very complex. The Heisenberg equations for the coupled oscillators, in the RWA ($g \ll \omega$), are 
\begin{equation}
  \frac{d}{dt} \left( \begin{array}{c} a \\  b \end{array} \right) 
   =   - \left( \begin{array}{cc}  \frac{\gamma}{2} &  i g \\  ig & \frac{\kappa}{2} \end{array} \right)   \left( \begin{array}{c} a \\  b \end{array} \right) +  \left( \begin{array}{l} \sqrt{\gamma} a_{\mbox{\scriptsize in}} \\  \sqrt{\kappa} [ b_{\mbox{\scriptsize in}} + \beta ]\end{array} \right)
\end{equation}
Here $\beta$ is complex, and gives the phase and amplitude of the signal driving the auxiliary resonator. The magnitude of $\beta$ is related to the microwave power driving the auxiliary by $|\beta | = \sqrt{P/(\hbar\Omega)}$~\cite{QNoise}. The symbols $a_{\mbox{\scriptsize in}}$ and $b_{\mbox{\scriptsize in}}$ are mutually independent quantum white-noise sources, characterized by the correlation functions~\cite{QNoise}
\begin{eqnarray}
    \langle a_{\mbox{\scriptsize in}} (t) a_{\mbox{\scriptsize in}}^\dagger (t+\tau) \rangle & = & (1 + n_T) \delta(\tau)  \\ 
        \langle a_{\mbox{\scriptsize in}}^\dagger (t) a_{\mbox{\scriptsize in}} (t+\tau) \rangle & = & n_T \delta(\tau) ,  
\end{eqnarray}
and $\langle b_{\mbox{\scriptsize in}} (t) b_{\mbox{\scriptsize in}}^\dagger (t+\tau) \rangle =  \delta(\tau)$. Solving these equation for $a(t)$ and $b(t)$, and setting $\gamma \ll \kappa$ and $g < \kappa/4$, the resulting steady-state coherent amplitude of the target resonator is  
\begin{equation}
   \langle a \rangle_{\mbox{\scriptsize ss}} = -i \beta \left( \frac{\sqrt{\kappa}}{g} \right) \left( 
   \frac{\kappa^2 - 6g^2}{\kappa^2 - 9g^2} \right) .  
\end{equation}
Thus the phase and amplitude of the target is controlled by the auxiliary. The rate at which the target responds to changes in the driving is given by the two decay constants $\lambda_{\pm} \approx \kappa/2 \pm \sqrt{\kappa^2/4 - 4 g^2} \sim \kappa$.  

The steady-state phonon number is $|\langle a \rangle_{\mbox{\scriptsize ss}} |^2$, plus the contribution $\langle n \rangle_{\mbox{\scriptsize c}}$ due to the temperature set by the cooling. While $ \langle n \rangle_{\mbox{\scriptsize c}}$ is approximately given by our simple expression in Eq.(\ref{eq::rsb}), the more exact expression, even within the RWA is rather complex:     
 \begin{equation}
 \langle n \rangle_{\mbox{\scriptsize c}} = \frac{\gamma n_T}{2} 
                                \left( \frac{\lambda_-^3}{N_-^4}  + \frac{4\lambda_-^2\lambda_+^2}{kN_-^2N_+^2} +  \frac{\lambda_+^3}{N_+^4}  \right)  , 
\end{equation}
where $N_\pm = (\lambda_\pm^2 + g^2)^{1/2}$. 

To summarize, we have shown that a frequency-converting, linear coupling between a high and low-frequency resonator can be used to cool the latter to its ground state using a state-swap. Further, in the steady-state, this process of state-swapping becomes a coherent feedback control loop, in which the mechanism of noise reduction corresponds to that of sideband cooling. This connection between state-swapping and coherent feedback control may provide insights into coherent control of more complex systems. 

\textit{Acknowledgements:} KJ is supported by the NSF under Project No.\ PHY-0902906, HN and MJ are supported by the Australian Research Council, and MJ is also supported by the Air Force Office of ScientiÞc Research (AFOSR) under grant AFOSR FA2386-09-1-4089 AOARD 094089. This work was performed with the supercomputing facilities in the school of Science and Mathematics at UMass Boston, as well as Prof.\ Daniel Steck's parallel cluster at the University of Oregon, which was funded by the NSF under Project No.\ PHY-0547926. 


%

\end{document}